\documentclass[5p]{elsarticle}
\journal{}

\makeatletter
\def\ps@pprintTitle{%
  \let\@oddhead\@empty
  \let\@evenhead\@empty
  \let\@oddfoot\@empty
  \let\@evenfoot\@oddfoot
}
\makeatother

\usepackage[T1]{fontenc}

\usepackage{bm}
\usepackage{lineno,hyperref}
\modulolinenumbers[2]

\usepackage{graphicx}
\usepackage[dvipsnames]{xcolor}

\usepackage{booktabs}
\usepackage{longtable}
\usepackage{graphicx}
\usepackage{dcolumn}
\usepackage{bm}
\usepackage{hyperref}
\hypersetup{
    colorlinks,
    linkcolor={blue},
    citecolor={blue},
    urlcolor={blue}
}
\usepackage{float}
\usepackage{color, colortbl}
\usepackage{changes}
\usepackage{amsmath,amssymb}
\usepackage{float}
\usepackage{xcolor}

\renewcommand{\vec}[1]{\mathbf{#1}}
\newcommand{\nvec}[1]{\hat{\mathbf{#1}}}

\definecolor{Gray}{gray}{0.9}
\usepackage{listings}
\definecolor{codegreen}{rgb}{0,0.4,0}
\definecolor{codegray}{rgb}{0.5,0.5,0.5}
\definecolor{codepurple}{rgb}{0.58,0,0.82}
\definecolor{backcolour}{rgb}{0.975,0.975,0.975}
\lstdefinestyle{mystyle}{
    backgroundcolor=\color{backcolour},   
    commentstyle=\color{codegreen},
    keywordstyle=\color{gray},
    numberstyle=\tiny\color{codegray},
    stringstyle=\color{codepurple},
    basicstyle=\ttfamily\footnotesize,
    breakatwhitespace=false,         
    breaklines=true,                 
    captionpos=b,                    
    keepspaces=true,                 
    numbers=left,                    
    numbersep=5pt,                  
    showspaces=false,                
    showstringspaces=false,
    showtabs=false,                  
    tabsize=2
}
\begin{document}

\title{
    Time Reversible Integration of the Landau-Lifshitz-Gilbert Equation
}
 
\author[UI,RWTH,FZJ]{Moritz Sallermann}
\ead{moritz.sallermann@rwth-aachen.de}
\author[UI]{Thorsteinn Freygardsson}
\author[UV]{Sergei Egorov}
\author[UI,LU]{Pavel Bessarab}
\author[UI]{Grzegorz Kwiatkowski}
\author[UI]{\ \ \ \ Hannes J\'{o}nsson}

\address[UI]{Science Institute and Faculty of Physical Sciences, University of Iceland VR-III,107 Reykjav\'{\i}k, Iceland}
\address[RWTH]{Department of Physics, RWTH Aachen University, 52056 Aachen, Germany}
\address[FZJ]{Peter Gr\"unberg Institut and Institute for Advanced Simulation, Forschungszentrum J\"ulich and JARA, 52425 J\"ulich, Germany}
\address[UV]{Department of Chemistry, University of Virginia,
Charlottesville,  VA 22901, USA}
\address[LU]{Department of Physics and Electrical Engineering, Linaeus University, Sweden}

\date{\today}

\begin{abstract}
A method for time-reversible numerical integration of the deterministic Landau-Lifshitz Gilbert equation by means of a second order Suzuki-Trotter decomposition is presented and tested against commonly used second order predictor-corrector methods. 
We find the time-reversibility of the Suzuki-Trotter integrator to be superior by several orders of magnitude 
while the
computational effort
is similar.
Calculations of trajectories backwards in time are useful, for example, when evaluating dynamical corrections to transition state theory.
\end{abstract}

\maketitle

\section{Introduction}
The dynamics of magnetic systems have been studied extensively and are an ever-growing field of research. Some of the earliest considerations of those kinds are the works of L. Landau and E. Lifshitz \cite{landau1992}, who proposed an equation of motion for magnetic moments in an effective field, which, today, is called the Landau-Lifshitz equation. Later T. L. Gilbert modified the equation by including a phenomenological damping term \cite{gilbert2004}, giving rise to the Landau-Lifshitz-Gilbert (LLG) equation. 

In modern day research, the importance of stable and accurate numerical solvers for the LLG equation cannot be overstated.
And while time reversibility is a more niche requirement, it can significantly shorten the computation times for calculations when dynamical corrections are applied for the rate coefficient from (harmonic) transition state theory \cite{bessarab2012}. A method, that is currently in use is called forward flux sampling, where a path between two minima on the energy surface is constructed and some interfaces are defined along the path. The method proceeds by finding the probability to reach each interface from the prior one via sampling of dynamical trajectories. The product of these probabilities, then, gives us the correction factor \cite{pellicelli2015}. This is computationally expensive since the probabilities are low to reach each interface are relatively low, as the trajectories are climbing the energy surface. If, instead, we could start at the dividing surface of the transition state and sample the trajectories forward and backward in time, it should give us a much cheaper way to estimate the correction factor. This has been done for polymer escape problems \cite{mokkonen2015} and should be applicable to magnetic systems as well.
In this context it time-reversibility of the used numerical solvers is highly desirable, but as it turns out this is not a feature of the commonly used predictor corrector methods utilized in modern computational frameworks for micromagnetism \cite{vansteenkiste2014,donahue1999} and atomistic spin dynamics \cite{skubic2008,muller2019}.
We implemented a time-reversible second order solver, making use of Suzuki-Trotter decompositions \cite{trotter1959, suzuki1976}, in the \textit{Spirit} code \cite{muller2019} 
and test the time-reversibility and accuracy of this proposed integrator against the SIB solver \cite{mentink2010}, the Depondt solver \cite{depondt2009}, the Heun solver \cite{sauer2014} and the fourth order Runge-Kutta method, implemented in the same framework.

\section{Model}
The deterministic Landau-Lifshitz Gilbert (LLG) equation describes the time evolution of classical spins in an effective magnetic field and is given by
\begin{align}
    \frac{\mathrm{d}\nvec{s}_i}{\mathrm{d}t} &= -\frac{\gamma}{1+\alpha^2} \nvec{s}_i \times \vec{B}_i - \frac{\gamma\alpha}{1+\alpha^2} \nvec{s}_i \times (\nvec{s}_i \times \vec{B}_i)\\
    &= \vec{F}_i(\nvec{s}_1 \dots \nvec{s}_N)
    \label{eq:LLG}
\end{align}
\noindent
where $\hat{s}_i$ represents a three dimensional unit spin vector for the $i$-th lattice site, $\gamma$ is the electron gyromagnetic ratio, $\alpha$ is the Gilbert damping parameter and $\vec{B}_i$ is the effective field acting on $\nvec{s}_i$. For ease of notation, we summarize the resulting time derivative into an effective term $\vec{F}_i$, which, generally, depends on all the aforementioned parameters and, via the effective field $\vec{B}_i$, on other spins in the system. The effective field is determined by taking the partial derivatives of an extended classical Heisenberg Hamiltonian $\mathcal{H}$ with respect to the spin directions 
\begin{equation}
    \vec{B}_i = \frac{\partial \mathcal{H}}{\partial \nvec{s}_i}.
    \label{eq:eff_field}
\end{equation}

A commonly used form of the Heisenberg Hamiltonian is
\begin{equation}
    \mathcal{H} = \mathcal{H}_\mathrm{exc} + \mathcal{H}_\mathrm{dm} + \mathcal{H}_\mathrm{an} + \mathcal{H}_\mathrm{ext},
    \label{eq:Hamiltonian}
\end{equation}
where the symmetric exchange interaction with strengths $J_{ij}$ is given by
\begin{equation}
    \mathcal{H}_\mathrm{exc} = -\sum_{\langle i,j\rangle} J_{ij}\nvec{s}_i\cdot\nvec{s}_j,
    \label{eq:hamiltonian_exc}
\end{equation}
and the Dzyaloshinskii-Moriya interaction (DMI) with interaction vectors $\vec{D}_{ij}$ is given by
\begin{equation}
    \mathcal{H}_\mathrm{dm} = \sum_{\langle i,j \rangle} \vec{D}_{ij} \cdot (\nvec{s}_i \times \nvec{s}_j),
\end{equation}
where the notation $\langle i,j \rangle$ signifies summation over \textit{unique pairs} of spins.
Furthermore, the uniaxial anisotropy with strength $K$ and axis $\nvec{k}$ has the form
\begin{equation}
    \mathcal{H}_\mathrm{an} = -\sum_i K (\nvec{s}_i \cdot \nvec{k})^2
    \label{eq:hamiltonian_anisotropy}
\end{equation}
and the Zeeman interaction of spins with magnetic moments $\mu_i$ in an external magnetic field $\vec{B}_\mathrm{ext}$ is denoted as 
\begin{equation}
    \sum_i \mu_i\hat{s}_i\cdot \vec{B}_\mathrm{ext}.
\end{equation}
We note that additional interactions, which may be of interest, include anisotropies with different symmetries (i.e. cubic), effective quadruplet and triplet interactions as well as the classical magnetostatic dipole-dipole interactions. 

\begin{figure*}[!t]
    \centering\includegraphics{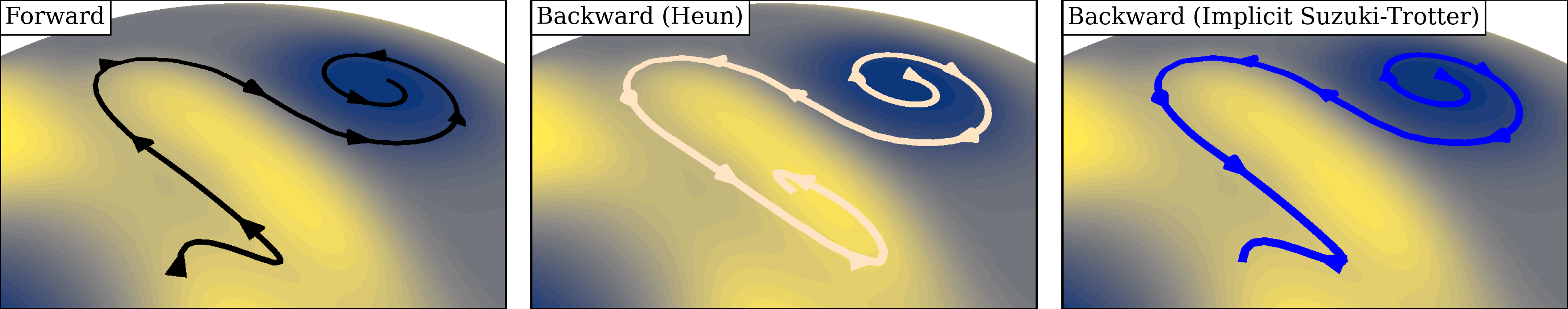}
    \caption{Illustration of forward and backward time trajectories computed with the Heun solver and the Suzuki-Trotter decomposition in a simple two dimensional energy landscape, composed of several contributions in the shape of anisotropic Gaussians. The two forward trajectories (black, left panel) in this illustration are indistinguishable. For the backward trajectories, a notable difference is found. The Heun solver (mid panel, beige) deviates from the forward trajectory and does not return to the initial configuration. The Suzuki-Trotter solver, on the other hand, reproduces the initial configuration correctly.}
    \label{fig:2d_illustration}
\end{figure*}


\section{Methods}
\subsection{Suzuki-Trotter Decomposition}
Assume a differential equation describing the time evolution of some state vector $S$,
\begin{equation}
    \frac{\mathrm{d}S}{\mathrm{d}t} = \left(\sum_{i=1}^N \mathcal{A}_i\right) S
    \label{eq:ST2}
\end{equation}
where $\mathcal{A}_i$ are operators acting on $S$, with the formal solution
\begin{equation}
    S(t+\Delta t) = \exp\left(\sum_{i=1}^N \mathcal{A}_i \Delta t\right)S(t).
    \label{eq:ST_formal_sol}
\end{equation}
In general, the operators $\mathcal{A}_i$ do not commute and, consequently, Eq.~\eqref{eq:ST_formal_sol} \emph{cannot} be decomposed into a product of single operator exponential functions $\prod_i \exp\left(\mathcal{A}_i\right)$. 
The second order Suzuki-Trotter (ST) decomposition approximates Eq.~\eqref{eq:ST_formal_sol} as
\begin{equation}
    S(t+\Delta t) \approx
    \prod_{j=N}^1 \exp\left(\mathcal{A}_{j} \frac{\Delta t}{2}\right) 
    \prod_{i=1}^N \exp\left(\mathcal{A}_i \frac{\Delta t}{2}\right)S(t),
    \label{eq:ST_second_order}
\end{equation}
with an error of $\mathcal{O}(\Delta t^3)$. Notice the reversed order in the second product of single operator exponentials, where the index starts at $j=N$ and ends at $j=1$.

The ST decomposition \eqref{eq:ST_second_order} can be applied to the LLG equation \eqref{eq:LLG} by defining the state vector $S$ as the composition of $N$ three-component spin vectors $\nvec{s}_i$
\begin{equation}
    S(t) = 
        \begin{pmatrix}
            \nvec{s}_1(t)\\
            \vdots\\
            \nvec{s}_N(t)
        \end{pmatrix},
\end{equation}
and by choosing the operators $\exp \left( \mathcal{A}_i \frac{\Delta t}{2} \right)$ to be the action of picking out spin $i$ and evolving it forward in time by $\Delta t / 2$ according to the LLG equation. Unless the spins are non-interacting, these operators do not commute.

\subsection{Single Spin Updates}
In order to apply the second order ST decomposition \eqref{eq:ST_second_order} in the aforementioned way, the system of spins is traversed in an arbitrary order and every visited spin is propagated by half a timestep $\Delta t/2$, while keeping all other spins fixed. Then, the spins are propagated a second time by another half a time step, this time reversing the previous order. It is crucial that the effective magnetic field $\vec{B}_i$, entering the LLG equation \eqref{eq:LLG}, is updated after each propagation of a single spin.\\
The second order ST decomposition ensures that the order of single spin updates is invariant under time reversal. To achieve a fully time reversible scheme, however, it is crucial that he discretised updates of single spins are time-reversible as well. In the following, we will achieve this by performing these updates using an implicit midpoint method. 
We note that this implicit midpoint method could also be used on the entirety of the spin system at once, without ST decomposition, to obtain a time-reversible solver. However, this would be immensely expensive, due to the amount of computations of the effective field necessary. The ST decomposition allows us to use the implicit midpoint scheme on one spin at a time and can, therefore, be understood as a "divide and conquer" way of achieving time-reversibility.

\subsubsection{Implicit Midpoint Propagator}
\label{sec:imp_midpoint}
We implement the discretised action of the single-spin propagation operators $\exp\left( \mathcal{A}_i \frac{\Delta t}{2} \right)$ from Eq.~(\ref{eq:ST_formal_sol}) in a time-reversible manner with the following update scheme,

\begin{align}
    \exp\left( \mathcal{A}_i \frac{\Delta t}{2} \right) S(t) = 
    \begin{pmatrix}
        \nvec{s}_1(t)\\
        \vdots\\
        \nvec{s}_i\left(t +  \frac{\Delta t}{2}\right)\\
        \vdots\\
        \nvec{s}_N(t)\\
    \end{pmatrix}
\end{align}

where the evolved spin $\nvec{s}_i\left(t + \frac{\Delta t}{2}\right)$ is found from the LLG equation~(\ref{eq:LLG}) with the effective term $\vec{F}_i$ taken at the midpoint between $\nvec{s}_i\left(t\right)$ and $\nvec{s}_i\left(t + \frac{\Delta t}{2}\right)$, while keeping all other spins $\nvec{s}_{j\neq i}(t)$ fixed:

\begin{align}
    \nvec{s}_i\left(t + \frac{\Delta t}{2}\right)\approx\nvec{s}_i(t) + \frac{\Delta t}{2} \vec{F}_i\left( \frac{\nvec{s}_i(t) + \nvec{s}_i\!\left(t+\frac{\Delta t}{2}\right)}{2}\right) 
    \label{eq:imp_midpoint}
\end{align}
We note that, since Eq.~(\ref{eq:imp_midpoint}) is invariant under the substitutions
\begin{align*}
    \Delta t &\rightarrow -\Delta t\\
    \nvec{s}_i\left(t + \frac{\Delta t}{2}\right) &\rightarrow  \nvec{s}_i\left(t \right)\\
    \nvec{s}_i\left(t \right) &\rightarrow \nvec{s}_i\left(t + \frac{\Delta t}{2}\right),
\end{align*}
the discretised propagation is time reversible. Further, the implicit midpoint structure inherently conserves the length of the spins $|\nvec{s}_i(t)|$.

If an uniaxial anisotropy term~(\ref{eq:hamiltonian_anisotropy}) is present, the effective field depends linearly on $\nvec{s}_i$
\begin{align}
    \vec{B}_i(\nvec{s}_i) = \vec{B}^{(0)}_i + \mathcal{K} \nvec{s}_i
    \label{eq:B_from_K}
\end{align}
with
\begin{equation*}
    \mathcal{K} 
    =
    K
    \begin{pmatrix}
        \hat{k}_x^2 & \hat{k}_x\hat{k}_y &  \hat{k}_0\hat{k}_z \\
        \hat{k}_y\hat{k}_x & \hat{k}_y^2 &  \hat{k}_y\hat{k}_z \\
        \hat{k}_z\hat{k}_x & \hat{k}_z\hat{k}_y &  \hat{k}_z^2
    \end{pmatrix}.
\end{equation*}
Using Eq.~(\ref{eq:B_from_K}), it is possible to solve Eq.~(\ref{eq:imp_midpoint}) via self-consistency iterations, without recomputing the entire effective field $\vec{B}_i(\nvec{s}_i)$.

If the system contains no uniaxial anisotropy term~(\ref{eq:hamiltonian_anisotropy}), the effective field does not depend on the currently updated spin and, therefore, Eq.~\eqref{eq:imp_midpoint} can be solved analytically via the use of the Cayley transform \cite{krishnaprasad2001}.
Notably, this propagator preserves the energy in a non-dissipative system, as is also demonstrated in the bottom panel of Fig~\ref{fig:spin_chain_conservation}, for the example of a Heisenberg spin chain.

We also implemented a different propagator, based on the analytical solution for an isolated spin in an external field, which is described in \ref{app:field_aligned}.

\subsection{Assessing the Accuracy of Solvers via Extrapolation}
It is important to verify how accurately the different solvers trace the analytical solution of the LLG Equation~(\ref{eq:LLG}), which, unfortunately, is almost never known. We can, however, still determine the accuracy by taking inspiration from the Richardson extrapolation method  \cite{richardson1997}.\\
Let us assume an analytical solution of the LLG equation is given by 
\begin{equation}
    \nvec{s}_i(t) \text{ with } t \in [0,T] ~\forall~ i \in [1,N],
\end{equation}
and denote the solution, via a discretised scheme with timestep $\Delta t$, by
\begin{equation}
    \nvec{s}_i(t_j | \Delta t) \text{ with } t_j = j \Delta t.
\end{equation}
Now, we try to compute the error $\varepsilon$ at the endpoint of the trajectory ($t=T$), while keeping only the term of lowest polynomial order
\begin{align}
    \nvec{s}_i(t_j | \Delta t)_\alpha &= \nvec{s}_i(t_j)_\alpha + \varepsilon_{i\alpha}\\
    &\approx \nvec{s}_i(t_j)_\alpha + k \Delta t^n + \mathcal{O}(\Delta t^{n+1}),
    \label{eq:error_polynomial}
\end{align}
where $\alpha \in \{x,y,z\}$, $n$ is the integer order of the numerical scheme and $k$ is a constant prefactor to the error term.

\begin{figure}[!t]
    \centering\includegraphics{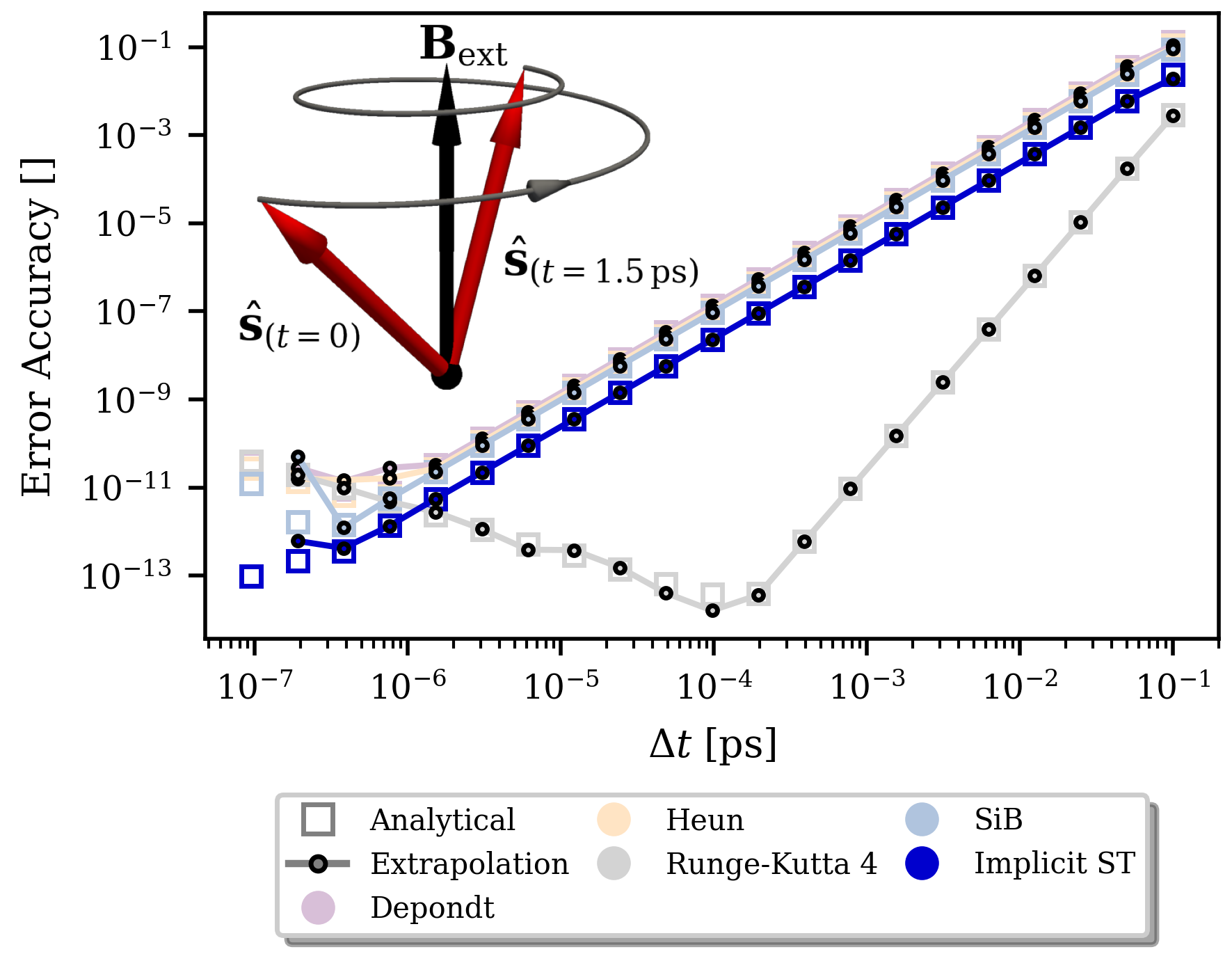}
    \caption{Accuracy over timestep for different solvers in the case of an isolated, damped spin in an external magnetic field. The field has been chosen such that the precession time is exactly $1\,\mathrm{ps}$. Different colors, correspond to different solvers. Continuous lines with dots denote errors found via the extrapolation method, while hollow squares denote errors found via the analytical solution of the trajectory. The inset illustrates the trajectory. The second order Suzuki-Trotter (blue) shows similar performance to the second order predictor-corrector schemes, while, unurprisingly, being worse than the fourth order Runge-Kutta solver (grey).}
\label{fig:single_spin_ext_field}
\end{figure}

By generating trajectories with three different timesteps ($2 \Delta t$, $\Delta t$ and $\Delta t / 2$) and neglecting all errors of higher order than $n$, we can solve for the three unknowns $k$, $n$ and $\nvec{s}_i(t)_\alpha$, resulting in 
\begin{align}
    n &= \log_2\left( \frac{\nvec{s}_i(T|2\Delta t)_\alpha - \nvec{s}_i(T|\Delta t)_\alpha}{ \nvec{s}_i(T|\Delta t)_\alpha - \nvec{s}_i\left(T| \Delta t /2 \right)_\alpha } \right)\\
    k &= \frac{1}{\Delta t^n}\frac{\nvec{s}_i(T|2\Delta t)_\alpha - \nvec{s}_i(T|\Delta 
    t)_\alpha}{2^n - 1}.
\end{align}
Consequently, we obtain an order $\mathcal{O}(\Delta t^{n+1})$ estimate for the maximum error over all spin components 
\begin{equation}
    \varepsilon(\Delta t) = \max_{i,\alpha} (\varepsilon_{i\alpha}) \approx \max_{i,\alpha} ( k \Delta t^n ) + \mathcal{O}(\Delta t^{n+1}).
    \label{eq:accuracy}
\end{equation}
In Fig.~\ref{fig:single_spin_ext_field} we verify this scheme by comparing it to exact errors, obtained via the analytical trajectory of a single spin in an external field, and observe that the errors estimates found by the extrapolation and the analytical trajectory are in good agreement.

The advantage of the proposed extrapolation method, in comparison to generating a reference trajectory with extremely low timesteps, is that it is (i) more time efficient and (ii) less sensitive to numerically induced errors occurring in update schemes with extremely low timesteps. A further application of the extrapolation method is finding optimal timesteps for a given system and integrator.

\section{Numerical Results}
The accuracy of the solvers in, perhaps, the simplest imaginable system, an isolated spin in a constant external magnetic field, is shown in Fig.~\ref{fig:single_spin_ext_field}. Here, it is observed that the three second order methods (Heun, SiB, Depondt and our Suzuki-Trotter decomposition) show very similar behaviour, meaning the error term is of the same polynomial order $n=2$, while differences stem from the constant prefactor $k$ (see Eq.~(\ref{eq:error_polynomial})). Accordingly, the fourth order Runge-Kutta solver shows much smaller errors with $n=4$ for larger timesteps, but has a minimum at around $\Delta t \approx 10^{-4}\,\mathrm{ps}$ and for lower timesteps the error increases again, due to finite precision in the floating point arithmetic, until it becomes worse than the second order solvers at $\Delta t \approx 10^{-6}\,\mathrm{ps}$. This is a general feature of numerical integration, where the optimal timestep is not known a priori and depends on details of the integration scheme as well as the system to be integrated. For timesteps between $10^{-6}$ and $10^{-7}\,\mathrm{ps}$ this increase in errors can also be observed for the second order methods.
In addition to this simple test system, we have chosen two more complex test systems. Parameters for all of the tested systems are to be found in \ref{app:test_systems}. 
In these more complex systems, we test (i) the accuracy of the solvers (in the sense of $\epsilon(\Delta t)$ from Eq.~(\ref{eq:accuracy})) and (ii) their time-reversibility, which measures how faithfully the solvers reproduce \emph{their own} trajectory, when reversing the direction of time (by flipping the sign of $\mathrm{d}t$ in Eq.~(\ref{eq:LLG})).

\begin{figure}[!t]
    \centering\includegraphics{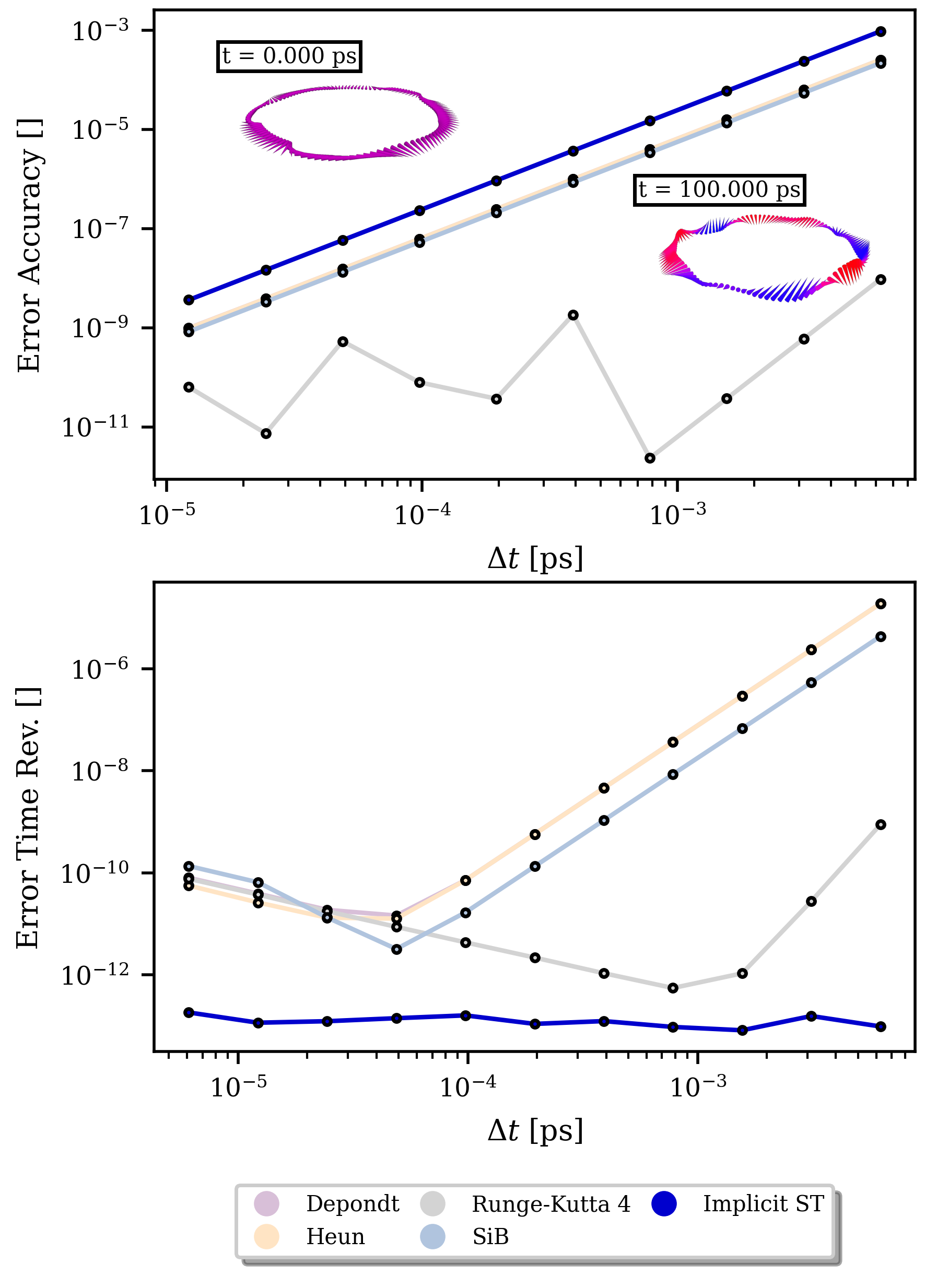}
    \caption{Forward accuracy (top) and time reversibility (bottom) for solvers in an undamped Heisenberg spin chain. The insets illustrate the initial configuration and the final configuration, after the time evolution has been computed for $100\,\mathrm{ps}$. Each arrow represents a spin and is colored according to the $z$-component of the spin direction. To represetn the periodic boundary conditions, the spin chain is illustrated as a ring.}
    \label{fig:spin_chain}
\end{figure}

\begin{figure}[!t]
    \centering\includegraphics{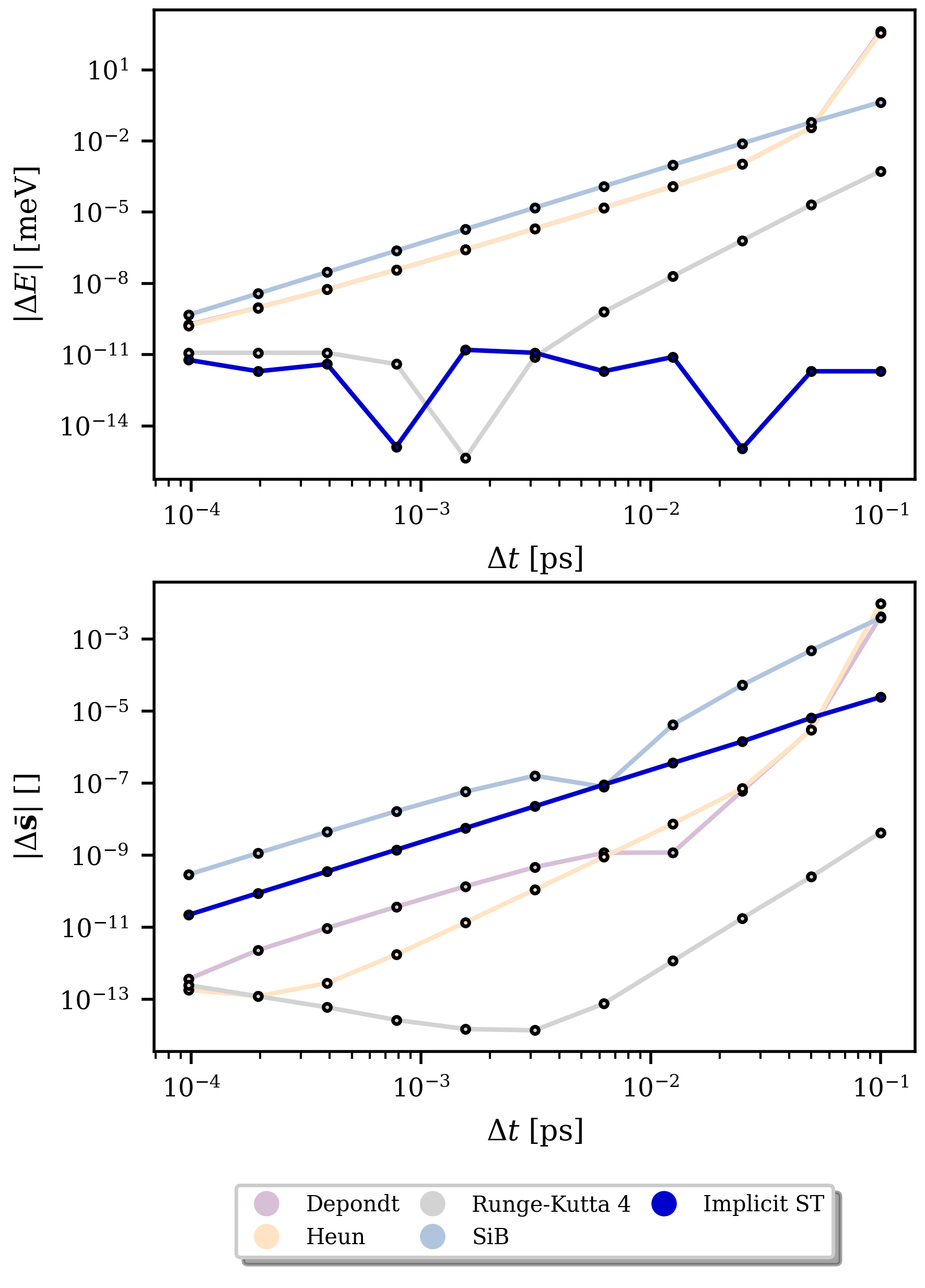}
    \caption{ Conservation of energy (top) and conservation of total magnetic moment (bottom) for solvers in the undamped Heisenberg spin chain.}
    \label{fig:spin_chain_conservation}
\end{figure}

The first system is a one dimensional Heisenberg spin chain with no damping ($\alpha = 0$ in Eq.~(\ref{eq:LLG})) and a Hamiltonian containing only isotropic exchange interactions (see Eq.~(\ref{eq:hamiltonian_exc})).
The top panel of Fig.~\ref{fig:spin_chain} shows the results for the forward accuracy, found from the aforementioned extrapolation method, since the analytical solution is not available here. Qualitatively, the 
results from Fig.~\ref{fig:single_spin_ext_field} are repeated, but with a different order of constant error prefactors $k$, which is not surprising since this quantity is, in general, system dependant.
The time reversibility is, comparatively, more interesting and depicted in the bottom panel. The reversibility of the Suzuki-Trotter solver is essentially independent of the time step and superior to all of the predictor corrector methods, which first show a decrease in timer reversal error, as the timestep is decreased but then increase again with even smaller timesteps, due to the accumulation of numerical errors.
Of further interest are two conservation laws, obeyed by this system. In Fig.~\ref{fig:spin_chain_conservation} the conservation of energy (top) and the conservation of average spin direction (bottom) are plotted. Here it is observed, that the implicit Suzuki-Trotter solver conserves the energy independently of the timestep and is, in this regard, superior to the other second order methods, while being superior to the fourth order Runge-Kutta solver for large timesteps.
The average spin direction is inherently conserved by neither of the solvers and hence the more accurate Runge-Kutta solver prevails. Interestingly, however, the second order predictor corrector methods show a more drastic violation of this conservation law as the timesteps become very large.

\begin{figure}[!t]
    \centering\includegraphics{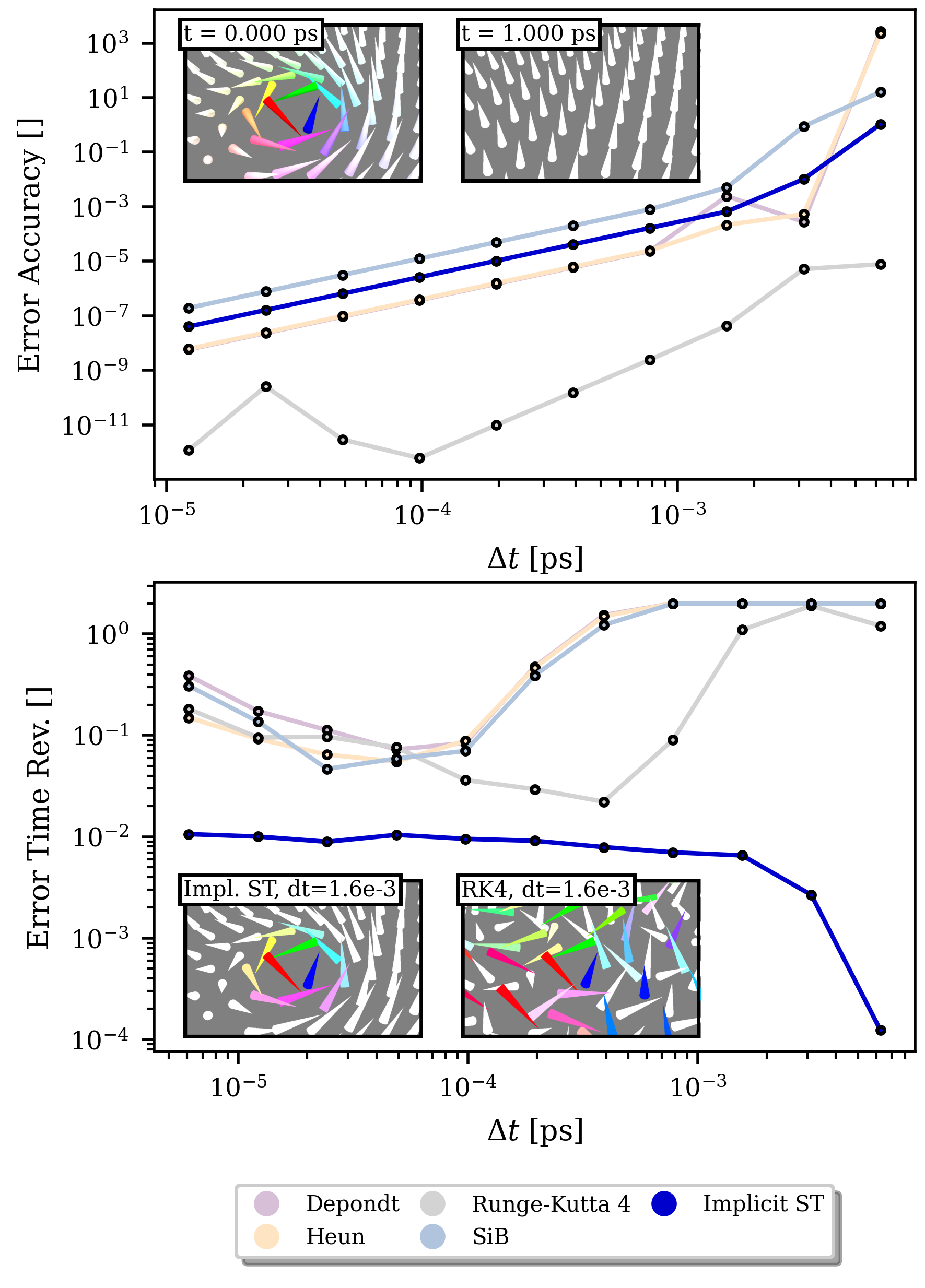}
    \caption{Forward accuracy (top) and time reversibility (bottom) for solvers in a damped chiral magnet, hosting a skyrmion. The insets, in the top panel, illustrate the initial configuration and the final configuration, after the time evolution has been computed for $1\,\mathrm{ps}$. Due to the non-zero damping the configuration, after the time evolution, is close to the ferromagnetic energy minimum. The insets, in the bottom panel, illustrate the resulting spin configuration when computing the time reversed trajectory, starting from the final configuration (right inset of the top panel). It is evident that the Suzuki-Trotter solver returns much more closely to the correct initial skyrmion, while the fourth order Runge-Kutta solver returns to a completely different configuration, due to its imperfect time-reversibility. }
    \label{fig:skyrmion}
\end{figure}

The second test system is a realistic model of a chiral magnetic fcc-Pd/Fe/Ir(111) monolayer, with interaction parameters determined from density functional theory calculations in Ref.~\cite{vonmalottki2017}, in which magnetic Skyrmions \cite{bogdanov1999} (vortex like topological solitons) can be stabilised. In contrast to the preceeding one dimensional spin chain, Gilbert damping is included. We initialize the system close to a first order saddle point for the transition of a skyrmion to the ferromagnetic ground state, so that the forward time evolution, due to the finite Gilbert damping, leads to the ferromagnetic state. Results for this setup are shown in Fig.~\ref{fig:skyrmion}.
Once again, the Suzuki-trotter solver shows comparable forward accuracy to the other second order integrators (top panel), while the fourth order method achieves the best results. With regards to time reversibility (bottom panel), the Suzuki trotter decomposition is superior to the other methods, but, notably, the achieved agreement between forward and backward trajectory is much lower than in the non-dissipative systems. The reason for this is that, due to the influence of damping, all trajectories contract towards local minima of the energy landscape (here, the ferromagnetic state) which in the asymptotic limit of infinitely long trajectories means that infinitely many initial configurations map to the same final configuration, hence making a reversal, inherently, impossible. But even if the trajectory is not reversed from from exactly a local minimum, but only close to it, the contraction of trajectories hinders an inversion of the reversal of the time evolution. Another observation is that, for large timesteps, the time reversibility of the Suzuki-Trotter solver improves. This tendency is explained by the fact that, if Eq.~(\ref{eq:imp_midpoint}) has a solution, it can be inverted and larger timesteps leads to less numerical computations and therefore less error accumulation. However, for too large timesteps the self-consistency iterations of Eq.~(\ref{eq:imp_midpoint}) do not converge in the first place.

\section{Discussion and Conclusions}
A time reversible method for the simulation of magnetic systems according to the deterministic LLG equation has been implemented.
The accuracy, when computing a trajectory forwards in time, is comparable to other well known second order solvers while the time-reversibility has been demonstrated to be much superior, in a non-exhaustive, yet comprehensive set of test cases.

Since each iteration of the Suzuki-Trotter solver needs two evaluations of the full gradient, if Eq.~(\ref{eq:B_from_K}) is used for the self-consistency iterations, the computational performance is, in principle, comparable to any of the second order predictor-corrector methods. In this proof of concept we have foregone a parallelized implementation, which is slightly harder to achieve, than for the predictor-corrector methods, as the sequential nature of the gradient computation leads to race conditions, if implemented naively. In cases, where only finite range interactions are of importance, like the exchange and DMI interaction, which decay rapidly as the distance between interaction sites increases, an easy way to achieve parallelization is to divide the simulation domain into cells which are larger than the effective interaction range. The Heisenberg Hamiltonian with only finite ranged interactions is widely used and especially suitable in quasi two-dimensional geometries.
Unfortunately the infinitely ranged dipole-dipole interaction may not always be negligible. This is especially problematic for three dimensional ferromagnetic systems. Here, a more sophisticated approach to efficiently evaluate the gradient of single spins, due to these dipole-dipole interactions, is called for. One possibility are tree based algorithms, like the Barnes-Hut \cite{barnes1986} or the Fast Multipole Method \cite{greengard1997}, which would allow these gradient computations with computational complexities of $\mathcal{O}(\log N)$ or $\mathcal{O}(1)$ \textit{per spin}, respectively ($N$ being the number of spins in the system).

\appendix
\section{Field aligned propagator}
\label{app:field_aligned}
For a single spin in an external field $\vec{B} = B\nvec{e}_z$, the analytical solution for the initial value problem posed by the LLG equation is known \cite{ma2011}:
\begin{align}
     \tan \frac{\theta(t+\Delta t)}{2} &= \tan \frac{\theta(t)}{2} \exp\left(- \frac{\gamma B \alpha \Delta t}{1 + \alpha^2}\right)\label{eq:analytical_theta}\\
    \phi(t+\Delta t) &= \phi(t) + \frac{\gamma}{1 + \alpha^2}B \Delta t\label{eq:analytical_phi},
\end{align}
where $\theta$ and $\phi$ denote the polar and azimuthal angle of the spin direction
\begin{equation*}
    \nvec{s}(t) = 
    \begin{pmatrix}
        \cos \phi(t) \sin \theta(t)\\
        \sin \phi(t) \sin \theta(t)\\
        \cos \theta(t) 
    \end{pmatrix}.
\end{equation*}

If no anisotropy is present, e.g. $\mathcal{H}_\mathrm{an} = 0$ in Eq.~\eqref{eq:Hamiltonian}, the update of single spins is mathematically equivalent to solving the LLG equation in a constant external magnetic field.
Therefore, one can make use of the known analytical solution Eqs.~(\ref{eq:analytical_theta}) and (\ref{eq:analytical_phi}) to perform the single spin updates, by intermediately transforming the spin vector into a local coordinate frame, in which the effective field points into the $z$-direction.
Care must be taken when including self interacting Hamiltonian terms of second order or higher in this method, since then the correct analytical solution is no longer given by Eqs.~(\ref{eq:analytical_phi}) and (\ref{eq:analytical_theta}). In this case the update is to be performed self-consistently and, similarly to the method described in Section~\ref{sec:imp_midpoint}, the effective field at the midpoint between $\nvec{s}(t)$ and $\nvec{s}(t+\Delta t))$ has to be used.


\section{Test systems}
\label{app:test_systems}
Parameters for the single spin in an external field, the Heisenberg spin chain and the chiral monlayer may be found in Tables~\ref{tab:single_spin}, \ref{tab:spin_chain} and \ref{tab:skyrmion}, respectively.

\begin{table}[!h]
\caption{Parameters for the system of a single spin in an external field of Figure~\ref{fig:single_spin_ext_field}.}
\label{tab:single_spin}
\centering
\begin{tabular}{lrl}
    \toprule
     $\vec{B}$ & $35.68\nvec{e}_z\,$T & External field \\
     $\alpha$  & 0.1 & Gilbert damping \\
      $N$ & 1 & Number of spins\\
     $T_\mathrm{final}$ & 1.5\,$\mathrm{ps}$ & Duration of trajectory\\
     \bottomrule
\end{tabular}
\end{table}

\begin{table}[!h]
\caption{Parameters for the Heisenberg spin chain of Figures~\ref{fig:spin_chain} and \ref{fig:spin_chain_conservation}.}
\label{tab:spin_chain}
\centering
\begin{tabular}{lrl}
    \toprule
     $J_1$ & 1$\,\mathrm{meV}$ & Exchange constant\\
     $\alpha$ & 0.0 & Gilbert damping\\ 
     $N$ & 128 & Number of spins\\
     $T_\mathrm{final}$ & 100\,$\mathrm{ps}$ & Duration of trajectory\\
     \bottomrule
\end{tabular}
\end{table}

\begin{table}[!h]
\caption{Parameters for the FCC-Pd/Fe/Ir(111)  monolayer of Figure~\ref{fig:skyrmion}.}
\label{tab:skyrmion}
\centering
\begin{tabular}{lrl}
    \toprule
    $\vec{B}$ &  $4\nvec{e}_z\,$T & External field \\
    $\nvec{K}$ & $\nvec{e}_z$ & Direction of uniaxial anisotropy\\
    $K$ & 0.7$\,\mathrm{meV}$ & Magnitude of uniaxial anisotropy\\
    $J_{1}$ & 14.4$\,\mathrm{meV}$ &     Exchange constant of 1st shell\\
    $J_{2}$ &  -2.48 $\,\mathrm{meV}$ &  " " " 2nd "\\
    $J_{3}$ &  -2.69 $\,\mathrm{meV}$ &  " " " 3rd "\\
    $J_{4}$ &  0.52 $\,\mathrm{meV}$ &   " " " 4th "\\
    $J_{5}$ &  0.74 $\,\mathrm{meV}$ &   " " " 5th "\\
    $J_{6}$ &  0.28 $\,\mathrm{meV}$ &   " " " 6th "\\
    $J_{7}$ &  0.16 $\,\mathrm{meV}$ &   " " " 7th "\\
    $J_{8}$ &  -0.57 $\,\mathrm{meV}$ &  " " " 8th "\\
    $J_{9}$ &  -0.21$\,\mathrm{meV}$ &   " " " 9th "\\
    $D_{1}$ & 1.00$\,\mathrm{meV}$ & DMI constant of 1st shell\\
    $\alpha$ & 0.1 & Gilbert damping\\ 
    $N$ & $64\cdot 64\cdot 1$ & Number of spins\\
    $T_\mathrm{final}$ & 1.00\,$\mathrm{ps}$ & Duration of trajectory\\ 
    Lattice & Triangular & Bravais lattice type\\
    \bottomrule
\end{tabular}
\end{table}

\section{Code listings}
C++ like pseudocode for the second order Suzuki-Trotter timestep can be found in Listing~\ref{lst:st}. An example implementation of the \texttt{evolve\_spin} function, using the implicit midpoint scheme, is contained in Listing~\ref{lst:imp_midpoint}.

\begin{lstlisting}[language=C++, caption=Second order Suzuki-Trotter timestep in C++ inspired pseudocode, label={lst:st}]
void suzuki_trotter(double dt, vector<Vec3> & spins)
{
    // N is the number of spins in the system
    for(int i=0; i<spins.size(); i++)
    {
        // Evolve i-th spin by half a timestep
        evolve_spin(i, dt/2.0, spins); 
    }
    // Update in reverse order
    for(int i=spins.size()-1; i>=0; i--)
    {
        evolve_spin(i, dt/2.0, spins);
    }
};
\end{lstlisting}

\begin{lstlisting}[language=C++, caption=Implicit midpoint propagator in C++ inspired pseudocode, label={lst:imp_midpoint}]
// Performs implicit midpoint update of a single spin
void evolve_spin(int i, double dt, vector<Vec3> & spins)
{
    // Convergence criterion
    const scalar convergence = 1e-16;
    // Max number of iterations
    const int max_iter       = 200;

    Vec3 spin_initial    = spins[i];
    Vec3 spin_previous   = spins[i];
    Vec3 spin_propagated = spins[i];
    Vec3 spin_avg        = spins[i];

    // force_callback should be implemented such that only the anisotropy contribution is re-computed

    Vec3 force_avg = force_callback(i, spins); // Force on i-th spin

    int iter = 0;
    bool run = true;

    while(run)
    {
        // Save the current spin
        spin_previous = spin_propagated;

        // Compute the propagated spin
        spin_propagated = spin_initial - dt*spin_avg.cross(force_avg);
        spin_propagated.normalize();

        // Compute the average spin
        spin_avg = 0.5 * (spin_propagated + spin_initial);
        spin_avg.normalize();

        // Compute the average force
        spins[ispin] = spin_avg.normalized();
        force_avg    = force_callback(ispin, spins);
        
        // change is the emaximum of thabsolute value of the componentwise difference between the two spins
        const scalar change = (spin_propagated - spin_previous).cwiseAbs().maxCoeff();

        iter++;
        run = change > convergence && iter < max_iter;
    }

    // Assign the propagated spin to the spins array
    spins[ispin] = spin_propagated;
}
\end{lstlisting}

\bibliography{Suzuki_Trotter_LLG.bib}

\end{document}